\documentstyle[11pt,aaspp4,epsf]{article}

\lefthead{Wandel}
\righthead{The Baldwin Effect in AGN: II}

\begin{document}


\def\myfig #1#2#3#4{\par
\epsfxsize=#1 cm
\moveright #2cm
\vbox{\epsfbox{#3}}
{\noindent Figure~#4 }\vskip .3cm }
\def\lg{\rm log}
\def\bh{black hole~}
\def\ew{equivalent width~}
\def\ad{accretion~disk~}
\def\el {emission-line}
\def\bhs{black~holes}
\def\ar{accretion~rate}
\def\ip{ionization-parameter~} 
\def\ed {Eddington~}
\def\ers{{\rm erg/sec}}
\def\ms{M_{\odot}}
\def\et{{\it et.al.}}
\def\sw{\rm Schwartzshild~}
\def\bb{black body~}
\def\be{Baldwin effect~}
\def\cmq{cm$^{-3}$}
\def\kms{km s$^{-1}$}


\title{On the Baldwin Effect in Active Galactic Nuclei:}
\title {II. Intrinsic Spectral Evolution}
\author{A. Wandel\altaffilmark{1}}
\affil{ Racah Institute, The Hebrew University, Jerusalem 91904, 
Israel}
\altaffiltext{1}{On leave at Astronomy Department, University of California,
    Los Angeles, CA 90095-1562}

\begin{abstract}

We suggest that
the growth of the central black hole in active galactic nuclei (AGN) 
due to the matter accreted over
the AGN  lifetime causes an evolution of the luminosity and spectrum.
In a previous paper  we have shown that the effective temperature 
of the UV continuum spectrum and the \el \ew are
anti-correlated with the black-hole mass. 
Here we estimate the change in the equivalent width of the emission 
lines due to the growth of the black hole mass caused by the accreted
material and
show that for several plausible accretion scenarios 
and effective-temperature models
the evolving equivalent width is anti-correlated with continuum luminosity,
implying that intrinsic evolution could  contribute to the Baldwin effect.

\end{abstract}

\keywords
 { galaxies: active --- galaxies: nuclei --- 
galaxies: Seyfert --- quasars: general ---
black hole physics --- accretion disks}

\section{INTRODUCTION}

Quasars and Active
Galactic Nuclei (AGN) are most probably powered by accretion of matter onto a supermassive black hole.
The amount of accreted matter required to explain the the observed energy
output of Seyfert galaxies and quasars is 0.001-10 solar masses per year, while the
central mass required to maintain steady accretion for the observed luminosity 
(the Eddington limit) is of the order of $10^6-10^9 \ms $.
Comparing these two numbers gives a timescale $M/\dot M\sim 10^7 - 10^9$ years, 
which indicates a  timescale for the evolution of the \bh  mass and
related properties, such as the emergent spectrum.

As the \bh  mass grows due to the accreted material, the observed luminosity and
spectrum  change with time. 
Since the continuum radiation ionizes the line-emitting material and
drives the broad AGN \el s, the change in the properties of the
radiation from the central source changes also the properties
of the emission lines. In particular the spectral shape of the ionizing 
radiation would determine the \ew   (the ratio between
the energy flux in the line to the continuum energy flux at the line 
frequency) of the \el s.
  
The accretion disk postulated in AGN provides a convenient 
(although model-dependent) framework for which
the emergent continuum spectrum may be calculated 
 (e.g. Sakura and Sunyaev, 1973), depending on the \bh  mass and the \ar .
Alternatively, an empirical average AGN UV spectrum may be related to the 
characteristic \bb temperature of the \ad . 

Baldwin (1977) has shown that the \ew of the CIV \el~ decreases with increasing 
continuum luminosity, a correlation known as the 'Baldwin effect'. A similar
relation has been found also for other broad \el s. The origin of this effect 
is not understood (for a recent review see Shields and Osmer, 1998).

In the first paper in this series (Wandel 1999a, hereafter paper I) 
we have argued that 
 the \be is related to the \bh mass. We have shown that the line \ew decreases with
increasing mass, so that the expected correlation between \bh mass and 
luminosity in luminous AGN implies an inverse correlation between \ew and luminosity.
 
In this work we suggest that at least part of the \be can be attributed to 
evolution. We estimate the relation between the \ew and the continuum spectrum, and
calculating the evolution of the continuum emission from the \ad 
for changing \bh mass and \ar we
show that for plausible evolutionary tracks the \ew is anti-correlated
with the continuum luminosity.

Section 2 gives an empirical relation between the model parameters 
($M$ and $\dot M$) and the implied evolutionary time scales,
as well as the {\it observed} spectrum and luminosity.  
In section 3 we suggest some plausible evolutionary scenarios for the \ar ,
and section in 4 we compose all these elements to produce the \be . 

\section{Accretion parameters and time scales}
\subsection{The Evolutionary timescale for accreting Black Holes}

In order to maintain steady spherical accretion the luminosity must be less than 
the \ed luminosity, 
$$L<L_{Edd}=4\pi GMm_pc/\sigma_T=1.3\times 10^{46}M_8 ~\ers
$$
or
\begin{equation}
\label {equ:med}
 M_8=0.7 \eta^{-1} L_{46}
\end{equation}

where $ M_8=M/10^8\ms$, $\eta=L/L_{Edd}$ is the \ed ratio, and $L_{46} =L/10^{46}$ \ers .

 We may define the
\ed time $t_E$ as the time required for an accretion-powered 
object radiating at the \ed luminosity to double its mass, 
due to accreted matter,
$$
t_E=Mc^2/L_E=4\times 10^8 {\rm y}.
$$
The observed luminosity implies an accretion rate of
$
\dot M = 0.16 \epsilon^{-1} L_{46} \ms {\rm y}^{-1}
$
where  $\epsilon$ is the efficiency.
Combining these two expressions gives the  mass, 
accumulated if the accretion rate is maintained during a time $t_E$ :
\begin{equation}
\label {equ:mev}
M_{8} = 0.6 \epsilon^{-1} (L/L_{Edd})^{-1}  L_{46} .
\end{equation}

The timescale for evolution (for example, significant change in the mass)
is given by
$$t_{ev}=M/\dot M=\epsilon \eta^{-1}T_E$$
for $\eta=0.1$ this gives $3\times 10^8$ yr for a \sw \bh and $1.6\times 10^9$yr
for a Kerr \bh .

\subsection{The Mass-Spectrum Relation}

The continuum \ad  spectrum
depends on the \bh mass. 
The spectrum of the \bb (and modified \bb ) disk is given by integrating over the entire
disk, 
\begin{equation}
\label {equ:fn}
L_\nu  \approx \int_{R_t}^{R_{out}} 2\pi R B_\nu [T(R) ] dR
\end{equation}

where $B_\nu(T)$ is the Planck function and
$R_t$ is the transition radius from the optically thick to optically thin
inner disk.

In paper I we have demonstrated that the basic property of the \ad spectrum -
thermal emission in the UV spectrum - leads to a peak in the SED, 
at an energy that is weakly decreasing with increasing \bh mass;
From the approximate shape of the optically thick \ad spectrum 
\begin{equation}
\label {equ:lker}
L_\nu  \approx A \left ({\nu\over \nu_{co}}\right )^{1/3} exp
\left (-{\nu\over \nu_{co}}\right )
\end{equation}
(where $A$ is a normalization constant and $\nu_{co}$ is the cutoff frequency)
it is possible to derive an expression for the peak (or cutoff) energy of the
UV spectrum:

\begin{equation}
\label {equ:ecocas}
E_{co}  \approx E_o M_8^{-1/2}L_{46}^{1/4}
\end{equation}
where $E_o\approx $17eV for an \ad around a \sw \bh and 
$E_o\approx$20eV for a disk around a maximally rotating Kerr \bh ,
and $E_o$ is related to the maximal temperature in the optically 
thick part of the disk by $E_o\approx 3kT_{max}$ for a \sw \bh
and $3.5kT$ for a Kerr one (Malkan 1990).

\subsection{Relating the Black Hole Mass and Accretion Rate to the UV spectrum}

Plotting the cutoff frequency versus the luminosity it is possible to infer
the mass and the \ar (or \ed ratio; fig. 1 in Paper I).
The $L/L_{Edd}$ ratio found in paper I is 0.01-1 for quasars, 
0.001-0.1 for Seyferts, while NLS1s are all in the
$L/L_{Edd}\approx 0.1-1$ range.

A similar analyses for the \sw \ad has been performed by Wandel and Petrosian 
(1988), who calculated for each pair of \ad 
parameters $M$ and $\dot M$  the actual observables, e.g. the UV luminosity and spectral index rather than fitting the blue-bump.
Wandel and Petrosian have calculated the \ad flux and spectral index 
 at 1450\AA , for a grid of \ad parameters $(M, \dot M)$. Inverting the grid they have obtained 
contours of constant \bh  mass 
 and constant \ed ratio ($\dot m$) in the $L-\alpha_{UV}$ plane (fig 2 ) .
 Plotting in this plane samples of AGN it is  possible to read off the contours the 
corresponding \bh mass and \ar . Comparing several groups of AGN a systematic trend 
appears: higher redshift and more luminous objects tend to have larger \bh  masses 
and luminosities closer to the \ed limit (see table~1). 
 Similar results were obtained by Sun and Malkan (1989).
\placetable {tbl1}
\begin{table}
\caption{Grouping of AGN \ad parameters.} \label{tbl1}
\begin{center}
\begin{tabular}{lccc}
\tableline
{}&{}&{}&{}\\
AGN group & Log$ F_{1450}$ & Log $M/\ms$ & $\dot M/\dot M_{Edd}$ \\ 
{}&{}&{}&{}\\
\tableline
{}&{}&{}&{}\\
Seyfert galaxies & 28-29.5 &7.5-8.5 & 0.01-0.5\\
Low Z quasars & 29-30.5 &8-9 & 0.02-0.1\\
Medium Z quasars & 30-31.5 & 8-9.5 & 0.1-0.5\\
High Z quasars & 31-32 & 9-9.5 & 0.03-2\\
{}&{}&{}&{}\\
\tableline
\end{tabular}
\end{center}
\end{table}

\myfig {12} 1 {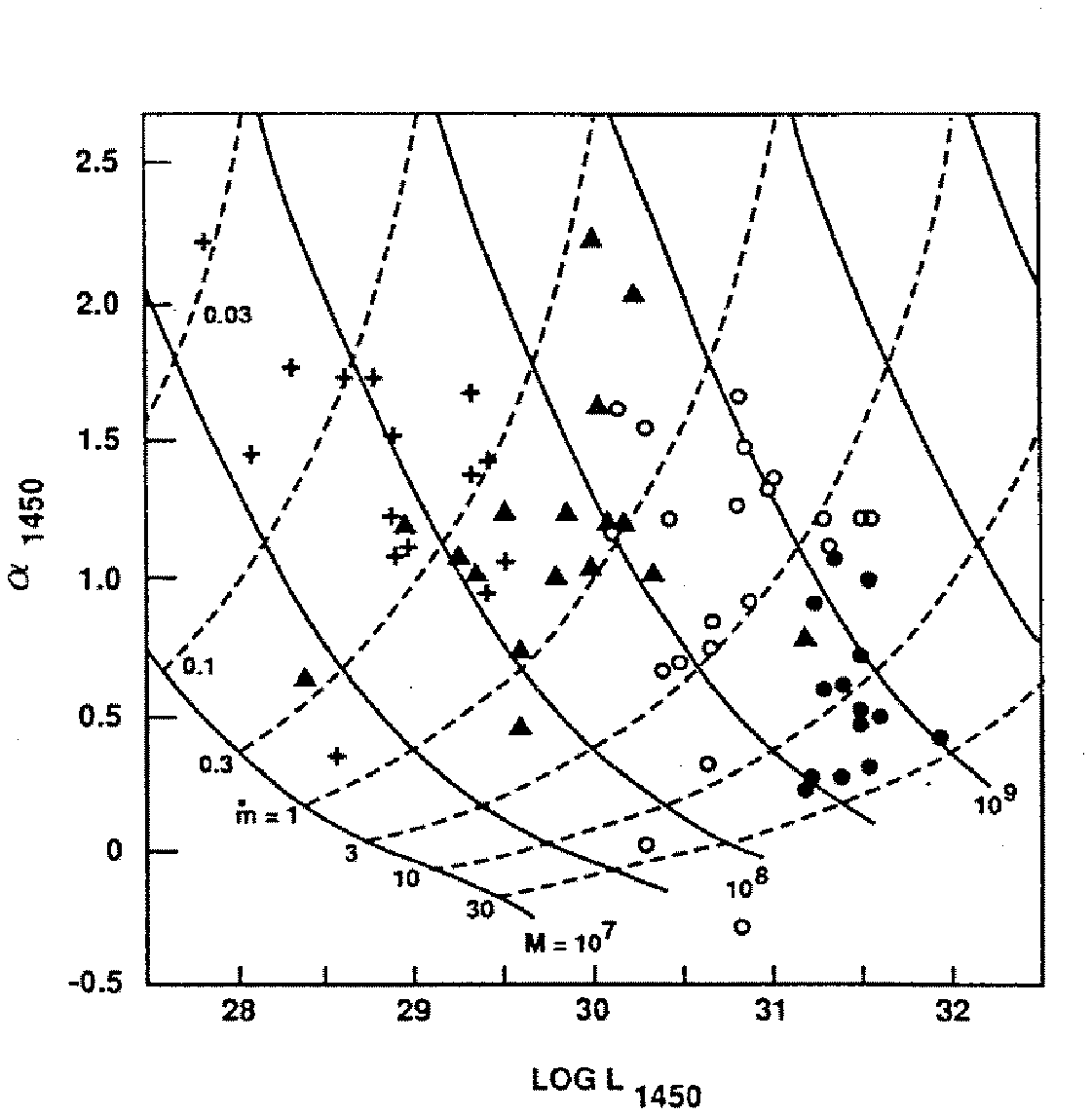}
{1. Accretion-disk parameters and spectral properties 
in the $\alpha -L$ plane (Wandel and Petrosian 1988). 
Crosses: Seyfert galaxies, triangles, open circles and filled circles:
low, medium and high redshift quasars, respectively. Continuous lines: constant mass;
dashed lines: constant $\dot M/M$ (note that  the labeling
of the curves in the figure follows the notation 
$\dot m=16.7\dot M/\dot M_{Edd}$).} 

\subsection {The Equivalent Width - Mass relation}

In Paper I we have shown that the line luminosity for some lines - in particular
CIV and OVI - is approximately proportional to the ionizing luminosity,
as the line emissivity is weakly dependent on the ionization parameter. The luminosity in the
line is given by

\begin{equation}
\label {equ:j}
L_{line}\approx L_{ion} f_{eff} j_0,
\end{equation}
where $f_{eff}$ is the effective covering factor,and $j_o$ is the peak emissivity.
The ionizing luminosity is determined by the maximal temperature of the
optically thick part of the \ad, which has been shown to depend on the \bh mass 
approximately as $M^{-1/4}$ for a variety of \ad models.
Combining these two results with eq. \ref{equ:ecocas} above gives the mass
dependence of the \ew - 
\begin{equation}
\label {equ:ewm}
 EW \propto exp[-(0.8-1.6)M_8^{1/4}]
\end{equation}
and similar dependencies for the luminosity ($EW\propto exp(C L_{46}^{1/4}$).
For the empirical power law AGN spectrum in the EUV ($\lambda < 1000\AA$)
we find $EW\propto L^{-\gamma}$ with $\gamma\sim 0.13-0.25$.

\section {Evolutionary Scenarios}
The evolution of the \bh  mass over cosmological times depends on the \ar , which may of
course change over time. From the theoretical efficiency and  observed luminosity we can estimate 
the present \ar . In order to determine the\bh  evolution, we need to specify the variation
of the \ar~ over time. We consider three representative scenarios:
\begin {itemize}
\item 
constant \ar
\item 
\ed limited accretion (constant \ed ratio)
\item 
spherical accretion from a homogeneous medium 
\end {itemize}
The tracks in the $E_{co}$ vs. $L$ plane of these three accretion scenarios 
are shown in fig. 2.
See also the discussion in Wandel (1999b).
\subsection {Constant \ar}
This would correspond to external feeding of the \bh , 
which is regulated by an external potential, e.g. by stellar encounters, 
where the stellar motions are governed by the gravitational potential in the bulge
of the host galaxy, far from the  influence of the \bh .
As the mass of the \bh  grows, $\dot M$ does not change, and hence
the accretion parameter, $\dot m\sim\dot M/M\sim M^{-1}$.
Eventually \ar~ will become very sub-\ed , and the \bh  will be starved,
or at least on a diet.
Since $L\sim \dot M$, in this scenario the (bolometric) luminosity also is also constant.
Note that the spectral energy distribution does change (though slowly) as $M$ increases.

\subsection {\ed limited accretion}
This would be the case in a \bh  that is over-fed.
The \ar~ cannot become very super-\ed , and the source will regulate the \ar~ to 
$L\approx L_{Edd}$ or
$\dot m\approx \epsilon^{-1}$.
Since $\dot m\sim \dot M/M$, $\dot m \sim$ constant implies
$\dot M\sim M$ and hence
$L\sim M$. 

\subsection {Spherical accretion from a homogeneous medium}
In this case we assume the matter supply comes from a homogeneous distribution 
(gas or stars), due to the gravitational potential of the \bh .
On large scales, the accretion will be spherical, and the accretion radius at which the \bh 
gravitational potential becomes significant is given by
\begin{equation}
\label {equ:rac}
R_{acc}\approx GMv_*^2 \approx 3 M_8v_{300}^2~ pc ,
\end{equation}
where
$v_*=300v_{300}$km/s is the stellar velocity dispersion.

The \ar~ is given by the Bondi formula
\begin{equation}
\label {equ:sar}
\dot M\approx 4\pi R_{acc}^2 {v_*}^2 \rho_* = (0.3 \ms /{\rm yr}) 
M_8^2v_{300}^{-3} \left ( {\rho_* \over 10\ms {\rm pc}^{-3}} \right ) ,
\end{equation}
where $\rho_*$ is the stellar number density.

In this case $L\sim\dot M\sim M^2$
and $\dot m\sim M$, that is, the \ed ratio increases with time.
Eventually the \ed ratio will approach unity and the accretion will become \ed limited.

\subsection {Accretion from an inhomogeneous medium}
A similar expression can be derived for a non-homogeneous medium, with a density profile
$\rho_*\sim R^{-p}$.
Assuming the velocity dispersion is independent of $R$,
eqs. \ref{equ:rac} and \ref{equ:sar} give for the \ar~ in that case
$$\dot M\sim M^{2-p}.$$
If also the velocity dispersion has a radial dependence,  $v_*\sim R^{-q}$
then we get a more complicated dependence, 
\begin{equation}
\label {equ:sarr}
\dot M\sim M^{{2\over 1+2q}-q-p}.
\end{equation}

In this case, any functional dependence is possible.
If for example $q=1/4$ and $ p=1/2$, 
$\dot M\sim M^{0.6}$,
and if  $q=1/2$ (point mass) and $ p=1/2$,
$\dot M =$const.

\myfig {17} 1 {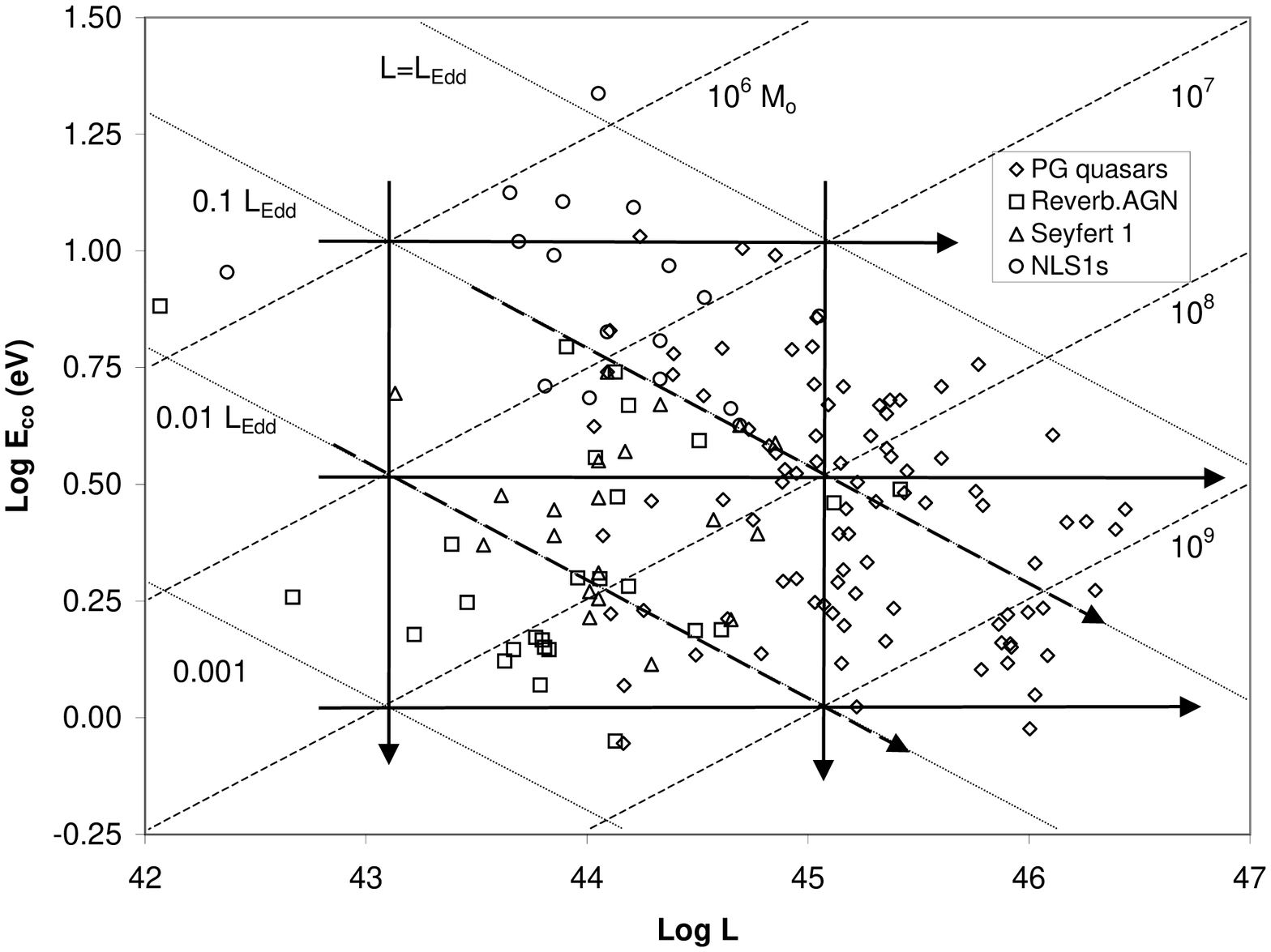}
{2. Black Hole evolution for the three \ar scenarios described in the text in
the cutoff energy vs. the monochromatic luminosity at
5100\AA. Diagonal dashed lines indicate constant \bh masses, and diagonal 
dotted lines indicate the \ed ratio.
Vertical arrows describe the direction of evolution for a constant \ar ,
Horizontal arrows show the evolution for a spherically accreting \bh from
a homogeneous medium, and diagonal dashed arrow are for accretion at
a fixed \ed ratio.
The superimposed data is
 PG quasars (diamonds; from Boroson and Green, 1994), Seyfert 1
galaxies (triangles and squares, the latter indicate AGN with BLR reverberation
data from Wandel, Peterson and Malkan 1999, with luminosities modified for
$H=50$) and Narrow Line Seyfert 1s  
(Boller, Brandt and Fink, 1996) }

\section{Putting All Together: an Evolutionary \be ?}

In the previous section we have found the relation between the accretion parameters 
$M$ and $\dot M$. It is easy to  find the functional dependence of the continuum luminosity on the
 \bh  mass for several accretion scenarios.
This cab be shown as evolutionary curves in the $M-\dot m$ and UV-spectrum vs. luminosity
 plane (figs. 1 and 2).
For example, an \ed bound \ad will evolve along the constant $\dot m$ curves
(dashed curves in fig. 1). 
A spherically accreting \bh  will evolve along curves of $\dot m\sim M$, which correspond
to nearly horizontal (from left to right) lines in fig. 1, and 
a constant \ar~ will yield $\dot m\sim M^{-1}$ which are almost vertical (upward) lines in fig. 1. Similar  evolutionary tracks are shown in fig. 2 (note 
however that the vertical axes of figs. 1 and 2 have an opposite sense).

Combining this with the dependence of the effective radiation-temperature
on the accretion parameters, and using the result that the \el~ luminosity is positively 
correlated with the radiation temperature, it is possible to find the relation between the 
equivalent width
and the continuum luminosity for each pair of accretion scenario and temperature model.

\placetable {tbl2}
\begin{table}
\caption{The relation between effective \ad temperature and \bh  mass for various 
accretion scenarios} \label{tbl2}
\begin{center}
\begin{tabular}{llll}
\tableline
{}&{}&{}&{}\\
Accretion scenario: &Constant $\dot M$& Constant $L/L_{Edd}$ & Spherical \\
$L(M)\propto\dot M$&const&$\sim M$&$\sim M^2$\\
\\
\tableline
$T_{eff}$ model:&{}&{}&{}\\
{}&{}&{}&{}\\
$T_{BB}$&$M^{-1/2}$ &$M^{-1/4}$&const\\
{}&{}&{}&{}\\
$T_{rad}$&$M^{-1/4}$ &$M^{-1/4}$&$M^{-1/4}$\\
{}&{}&{}&{}\\
$T(\tau_*=1)$ or $T$(BB/rad)&const &$M^{-1/4}$&$M^{-1/2}$\\
{}&{}&{}&{}\\

{}&{}&{}&{}\\
$T$(BB-var)$\propto L^{-1/4}$&const &$M^{-1/4}$&$M^{-1/2}$\\
{}&{}&{}&{}\\
\tableline
\end{tabular}
\end{center}
\end{table}

\subsection {The L-T relation}

The dependence of the effective temperature on the \bh  mass is  summarized in table 2. 
The header shows the three accretion scenarios and the 
corresponding dependence of the continuum luminosity on the mass.
The next row gives the dependence of the luminosity on the mass for each scenario.
The left column shows the temperature models. For each pair the table gives 
 the functional dependence of $T_{eff}$ on M.

Since the \bb -radiation boundary  and the photospheric models, have a very 
similar dependence on the accretion parameters, we combine them in one
row in tables 2 and 3. Table 2 has nine ($T_{eff},\dot M$) pairs,
for \ad models, and an additional row for the model independent \bb - variability 
estimate of $T_{eff}$ (marked ``$T$(BB-var)''). 

For all pairs except two  (constant $\dot M$ with $T({\rm BB-rad})$ and homogeneous medium 
accretion with $T_{BB}$).  The effective disk temperature is decreasing with increasing \bh  mass.
We note that this is true also for the general temperature-luminosity relation
 (eqs. 21,22 in Paper I), so  the \ad model is not essential.
In that case the anti-correlation between temperature and luminosity is built into the
the expression for the temperature.

For all accretion scenarios  the optical and UV continuum luminosity is increasing with $M$. 
This is certainly the case for \ed bounded accretion and for accretion from a homogeneous medium;
For a constant \ar~ the bolometric luminosity is constant, but as the \bh  mass
increases the peak of the \ad spectrum moves to lower frequencies. Since the optical or 
near UV continuum are on the Rayleigh-Jeans part of the \bb spectrum,
the luminosity in these bands increases as the mass increases even if 
the total luminosity remains constant.
However, since for a radially extended \ad the low energy slope in the near UV is much flatter
than for a Planck spectrum, this is a very small effect. As the $\dot M$=const is a limiting
case, and in reality the \ar~ is expected to grow, even if slower than linearly, as argued above,
we replace the $\dot M$=const. column in table 3 by a more general case,
$\dot M\propto M^\beta$, $\beta<1$, which allows for an evolution in luminosity.

Applying the dependence of L on M it is possible to deduce the dependence of T on L.
Relating the EW to the spectral UV peak energy (Paper I) we 
can quantify the expected \be for each of the models.

{\bf Decreasing \ar .}
In the three cases considered above, the \ar~ increases with time or remains constant.
What if the accretion rate decreases with time? 
In the two latter \ad -temperature models in Paper I (5.2.c and d) we find
$E_{max}\sim \dot M^{-1/3}$, so $T_{eff}$ increases, and so does the line luminosity.
If we take $L\propto \dot M$ the continuum luminosity decreases, and 
there is still a negative correlation.
This is the case also in the general L-T relation. 
In the  model, the temperature is independent of the
\ar , and since the mass increases, the 
For the \bb and radiation-pressure dominated models  the
situation is less clear, because $T_{eff}$  decreases with decreasing $\dot M$ 
(as $\dot M^{1/4}/M^{1/2}$ for the \bb model
and being independent of the \ar in the radiation-pressure dominated model, 
it still decreases as $M^{-1/4}$). Since $L\sim \dot M$ decreases, the question whether 
$T_{eff}$ is correlated or anti correlated with $L$ would depend on the details of the
behavior of the \ar .

\subsection {The \ew }
How does the \ew vary? The \ew is defined as $L_{line}/F_\lambda$,
where $F_\lambda$ is the flux per unit wavelength at the wavelength of the line.
In Paper I we have shown that $L_{line}$ depends on the ionizing continuum, which correlates 
strongly  with $T_{eff}$. The luminosity in the line is also roughly 
linearly dependent on the continuum luminosity, $L\sim \lambda F_\lambda$, so that
$$L_{line}= f( T_{eff}) L ,$$
$f(T)$ being a monotonic function with an exponential increase near 
$3kT\sim 1$Ryd.
This gives for the \ew 
$$EW(L)\sim L_{line}/L=  f[T(M)].$$

Applying the observed UV spectral slope, $L_\nu\propto \nu^{-\alpha}$ ($1.5<\alpha<2$),
we can derive from the Temperature -- Mass dependence shown in table 2 a relation between
\ew and luminosity.
From Paper I we have
$$EW\propto \left ( {E_{eff}\over E_{ion}}
\right ) ^{-(\alpha-1)}$$
so we can derive the \be expected for each model quantitatively.
Table 3 assumes $\alpha=2$.
For a $L_\nu\sim \nu^{-\alpha}~(\alpha< 2)$ spectrum the exponent of L
should be multiplied by $\alpha -1$.

\placetable {tbl3}
\begin{table}
\caption{The predicted \be (dependence of the EW on $L$)
for different effective \ad temperature models and 
accretion scenarios. An EUV spectrum of $L_\nu\sim \nu^{-2}$ has been assumed.
For a $L_\nu\sim \nu^{-\alpha}~(\alpha\neq 2)$ spectrum the exponent of L
should be multiplied by $\alpha -1$. $\dot m_{-1}=\dot m/0.1$. } 
\label{tbl3}
\begin{center}
\begin{tabular}{lllll}
\tableline
{}&{}&{}&{}&{}\\
Accretion scenario: &&$L\sim M^\beta$& $L/L_{Edd}$ & Spherical \\
$\dot m:$&&$\dot m\sim M^{\beta-1}$&const&$\sim M$\\
\\
\tableline
$T_{eff}$ model:&$kT(L,\dot m)$(eV)&{}&{}&{}\\
{}&{}&{}&{}\\
$T_{BB}$&$4\dot m^{1/2}_{-1}L_{46}^{-1/4}$&$L^{(2\beta-3)/4}M^{(1-\beta)/2}$ &$L^{-1/4}$&const\\
{}&{}&{}&{}&{}\\
$T_{rad}$&$4\dot m^{1/4}_{-1}L_{46}^{-1/4}$ &$L^{(\beta-2)/4}M^{(1-\beta)/4}$&$L^{-1/4}$&$L^{-1/8}$\\
{}&{}&{}&{}&{}\\
$T(\tau_*=1)$\tablenotemark{a} &$3 L_{46}^{-1/4}$&$L^{-1/4}$ &$L^{-1/4}$&$L^{-1/2}$\\
{}&{}&{}&{}&{}\\

{}&{}&{}&{}&{}\\
$T$(BB-var)\tablenotemark{b}&$10L_{46}^{-1/4}$&$L^{-1/4}$ &$L^{-1/4}$&$L^{-1/4}$\\
{}&{}&{}&{}&{}\\
\tableline
\end{tabular}
\tablenotetext{a}{and the $T(BB/rad)$ boundary model}
\tablenotetext{b}{$T\propto L^{-1/4}$}
\end{center}
\end{table}

We see from table 3 that for most combinations of effective temperature model and accretion
scenario there is an evolutionary \be , and the predicted slope ($dlnEW/dlnL$) is 
$\gamma = 0.25(\alpha -1)$, consistent with the observed \be of 0.15-0.2.


\subsection {The Exceptions}
We conclude that for many accretion-scenarios and temperature-models
the line luminosity decreases with increasing mass, except
the two for which the effective temperature is constant, implying an evolutionary
\be : as the \bh  evolves over cosmological times due to the mass accreted, the continuum
luminosity (optical or near UV) increases, while the \ew decreases.
Also the two pairs with a constant effective temperature will eventually show a \be .
To see this, note that for the ($T_{BB}-\dot M$(spherical)) pair, $\dot m$ is increasing
as $M$, so it will eventually approach the \ed \ar , and move from the $T_{BB}$ model
to one of the other $T_{eff}$ models,  which (for $\dot M$(spherical)) do have a \be .
In the other pair, (constant $\dot M$ with $T({\rm BB-rad})$), the situation is opposite,
but with the same result: since there $\dot m\sim M^{-1}$, eventually the \ar~ will become
enough sub-\ed so that the $T_{BB}$ model will apply, which for the constant \ar~
scenario does have a \be .

\section{discussion}

In paper I  we have made the case for a \be due to the dependence 
of the \ew on the \bh mass. We have shown that the line \ew decreases with
increasing mass, so that the expected correlation between \bh mass and 
luminosity in luminous AGN implies a \be .

In this work we show that an intrinsic evolution of the \bh mass and \ar
can create an evolutionary \be .
In addition, the 
relation between the emission-line luminosity and the
ionizing continuum shape depends on the unknown density profile
of the line emitting gas, and on the unknown differential covering factor,
which could show intrinsic evolution (for an individual object along it's 
lifetime) as well as cosmological evolution.

 How can we infer whether evolution is the {\it main cause}
of the Baldwin effect?  This work shows that
evolution is a possible cause.  But there are several
other possible causes as well.  For example, the ranges
in actual black hole masses and accretion rates in
individual AGN might deviate greatly (and systematically)
from the average relations assumed. 

\section{Summary}
We suggest that
the growth of the central \bh  mass due to the  accreted matter
causes  the luminosity and the spectrum to
evolve over over the AGN life time.
Using  general model-independent relations, or the thin \ad  spectrum,
we estimate the evolution of the \ew and the continuum spectrum, and
show that for plausible evolutionary tracks 
as well as for the model-independent \bb temperature estimate and for
most variants of the thin \ad model
the \ew decreases with increasing continuum luminosity
 implying an evolutionary origin to the Baldwin effect.
 
\acknowledgments
 
Stimulating discussions with Jack Baldwin during
the meeting "Quasars as Standard Candles in Cosmology" at La Serena, 
Chile, triggered this work.
I also acknowledge valuable discussions with Gary Ferland, Matt Malkan
and Vah\' e Petrosian, the contribution of an anonymous referee,
and the hospitality of the Astronomy Department at UCLA and 
Stanford University.


\begin{references}
\reference   {}
Baldwin, J.A.  1977, \apj , 214, 679.
\reference {} Boller, Th., Brandt, N. and Fink, V. 1996, A\& A 305, 53
\reference {} Boroson, T. and Green, R.F. 1992, ApJS 80, 109

\reference  {}Edelson, R. and Nandra, K. 1999, ApJ
\reference   {}Malkan, M.A. 1990, in IAU Colloquium no. 129,
"Structure and Emission Properties of Accretion Disks", 
eds. C. Bertout et.al.,  Editions Frontiers: Paris, p. 165.
\reference  { }
Shields, J. and Osmer, P. 1998 in "Quasars as Standard Candles for Cosmology",
La Serena, Chile, ed. G. Ferland, ASP.
\reference  { }
Sun, W.H. and Malkan, M.A. 1989, \apj , 346, 68.
\reference  { }
Wandel, A. 1997, \apjl 430, 131.
\reference  { }Wandel, A. 1999a (Paper I), \apj, in press 
\reference  { }Wandel, A. 1999b, \apjl, 509, L39
\reference  { } 
Wandel, A. and Mushotzky, R.F. 1986, \apjl , 306, L61.
\reference  { }Wandel, A., Peterson, B.M. and Malkan, M.A. 1999, ApJ, 526,
\reference  { }
Wandel, A. and Petrosian, V. 1988, \apjl , 329, L11.
\end{references}
\end{document}